# Propagation of Spin-Polarized Electrons Through Interfaces Separating Differently Doped Semiconductor Regions

Yuriy V. Pershin and Vladimir Privman
Center for Quantum Device Technology, Department of Electrical and Computer Engineering and
Department of Physics, Clarkson University, Potsdam, New York 13699-5720, USA
pershin@clarkson.edu, privman@clarkson.edu

*Abstract.* High degree of electron spin polarization is of crucial importance in operation of spintronic devices. We study the propagation of spin-polarized electrons through a boundary between two n-type semiconductor regions with different doping levels. We assume that inhomogeneous spin polarization is created/injected locally and driven through the boundary by the electric field. The electric field distribution and spin polarization distribution are calculated within a two-component drift-diffusion transport model. We show that an initially created narrow region of spin polarization can be further compressed and amplified near the boundary. Since the boundary involves variation of doping but no real interface between two semiconductor materials, no significant spin-polarization loss is expected. The proposed mechanism will be therefore useful in designing new spintronic devices.

## I. INTRODUCTION

Recent proposals for devices based on the manipulation of electron spin [1-11] have inspired a renewed interest in theoretical and experimental investigations of spin-related effects in semiconductors [12-16]. Operation of a spintronic device requires efficient spin injection into a semiconductor, spin manipulation, control and transport, and also spin detection. Once injected into a spintronic device, electrons experience spin-dependent interactions with the environment, which cause relaxation. Electron spin polarization is lost at the interfaces between two semiconductor materials as well. It is important to understand the mechanisms of electron spin relaxation and find the ways to increase electron spin polarization density.

In the present work we study the motion of spin-polarized electrons injected near a boundary between two n-type semiconductor regions having different properties, e.g., doping level, mobility, electron spin relaxation time; see Figure 1. Two-component (spin-up and spin-down) drift-diffusion model

This research was supported by the National Science Foundation, grants DMR-0121146 and ECS-0102500, and by the National Security Agency and Advanced Research and Development Activity under Army Research Office contract DAAD 19-02-1-0035.

in an applied electric field is used. We assume that localized spin polarization is created and driven through the boundary by the electric field. Two types of spin-polarization source are considered: instantaneous source and continuous one. We take into account charge accumulation/redistribution near the boundary. The latter effect is analogous to the depletion region formation in a p-n [17] junction, and it introduces coordinate-dependent electric field in the equation for the spin polarization density. We solve the resulting differential equations for the electric field and spin polarization density.

We have obtained an interesting result concerning propagation of spin-polarized current through a boundary between two semiconductor regions with different doping levels. It was found that the spin polarization density can be condensed and amplified near the boundary. The built-in electric field at the boundary accelerates propagation of the spin polarization through the boundary, if spin polarization passes from the low-doped region to high-doped region. Spin amplification occurs past the boundary, within the distance of the order of the depletion layer width. We point out that this mechanism, involving only the doping variation, has the advantage of not requiring a materials interface, thus avoiding additional spin-polarization losses [18]. The details of our model and calculations can be found in [19].

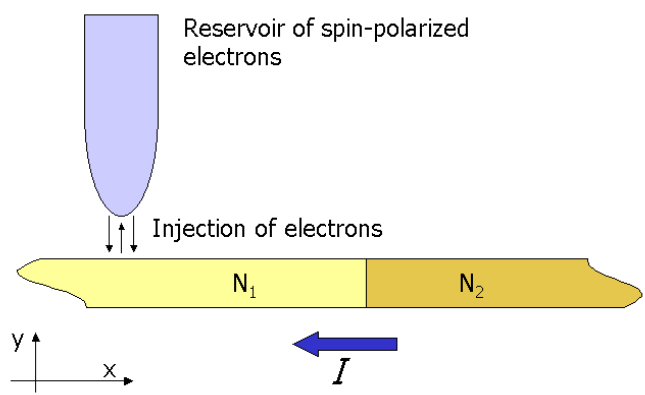

**Figure 1.** Schematic representation of the system under study: spin-polarized electrons are injected into the low-doped region.

## II. SYSTEM UNDER STUDY

We consider the n/n$^+$ semiconductor junction, Figure 1, with the low-doped region n for $x < 0$, and the high-doped region n$^+$ for $x > 0$. The corresponding donor densities are $N_1$ and $N_2$. We are not specifying the length of these regions, selecting the current as the external control parameter rather than the applied voltage. Moreover, the structure is assumed to be sufficiently thick in transverse directions to allow for one-dimensional electrostatic treatment.

We assume that a localized source of spin-polarized electrons is located in the low-doped region. The injection from the source results in a difference in concentrations of electrons with opposite spin direction. Under influence of the electric field, the non-equilibrium spin polarization drifts in the direction of the high-doped region.

From the experimental standpoint, non-equilibrium spin-polarization can be created locally in the bulk of semiconductor, for example, by using ferromagnetic-metal scanning tunneling microscopy (STM) tips [20,21], or by optial pumping techniques [22-24]. Alternative spin polarization mechanisms are also possible [25-27]. In following sections, we study the propagation of non-equilibrium spin polarization through the boundary between the low-doped and high-doped regions.

## III. PHYSICAL MODEL

We use the two-component drift-diffusion model [28,29] to describe the evolution of non-equilibrium electron spin polarization. The following set of equations for spin-up and spin-down electrons is used,

$$e\frac{\partial n_{\uparrow(\downarrow)}}{\partial t} = \text{div}\, \vec{j}_{\uparrow(\downarrow)} + \frac{e}{2\tau_{sf}}\left(n_{\downarrow(\uparrow)} - n_{\uparrow(\downarrow)}\right) + S_{\uparrow(\downarrow)}(\vec{r},t), \quad (1)$$

$$\vec{j}_{\uparrow(\downarrow)} = \sigma_{\uparrow(\downarrow)}\vec{E} + eD\nabla n_{\uparrow(\downarrow)}, \quad (2)$$

$$\text{div}\, \vec{E} = \frac{e}{\varepsilon\varepsilon_0}(N_i - n), \quad (3)$$

where $-e$ is the electron charge, $n_{\uparrow(\downarrow)}$ is the density of spin-up (spin-down) electrons, $j_{\uparrow(\downarrow)}$ is the current density, $\tau_{sf}$ is the spin relaxation time, $S_{\uparrow(\downarrow)}$ describes the source of the spin polarization, $\sigma_{\uparrow(\downarrow)} = en_{\uparrow(\downarrow)}\mu$ is the conductivity, with $n = n_\uparrow + n_\downarrow$ the electron density, and $\mu$ the mobility, connected with the diffusion coefficient $D$ via the Einstein relation $\mu = De/(k_BT)$, and defined via $\vec{v}_{drift} = \mu\vec{E}$. Equation (1) expresses the continuity equations for spin-up and spin-down electrons including spin-flip processes, Eq. (2) is the expression for the current density which consists of the drift current and the diffusion one, and Eq. (3) is the Poisson equation. Combining Eqs. (1)-(3), we obtain the second-order nonlinear equation for the steady-state electric field profile,

$$\frac{\partial^2 E}{\partial x^2} + \frac{e}{kT}E\frac{\partial E}{\partial x} - \frac{e^2 N_i}{kT\varepsilon\varepsilon_0}E = -\frac{j_0}{\varepsilon\varepsilon_0 D} + \frac{e}{\varepsilon\varepsilon_0}\nabla N_i, \quad (4)$$

and the equation for the spin polarization density $P = n_\uparrow - n_\downarrow$,

$$\frac{\partial P}{\partial t} = D\Delta P + D\frac{e\vec{E}}{k_BT}\nabla P + D\frac{e\nabla\vec{E}}{k_BT}P - \frac{P}{\tau_{sf}} + F(\vec{r},t). \quad (5)$$

Here $F(r,t) = [S_\uparrow - S_\downarrow]/e$ represents a spin polarization density created by the external source. The spin polarization density is coupled to the charge density through the electric field. Thus, our numerical calculation involves two steps: first, the electric field profile is found as the solution of Eq. (4) and, second, Eq. (5) is solved for the spin polarization density.

## IV. RESULTS AND DISCUSSION

We have solved Eqs. (4)-(5) numerically, for different values of the parameters [19]. The obtained spin polarization density profiles are qualitatively similar. The results presented here, calculated for a selected typical set of parameter values, are representative of the general idea of our proposal for spin-polarization amplification. The obtained stationary electric-field profile is shown in Figure 2 (the blue line). The charge is redistributed near the boundary to equilibrate the Fermi levels of the regions with different doping, forming the dipole layers with a positive charge at the $n^+$ region, and a negative charge at the $n$ region. This dipole produces the spike of the electric field at the boundary, which extends into the both regions. This spike produces a voltage drop (built-in voltage). Away from the boundary, the electric field is constant.

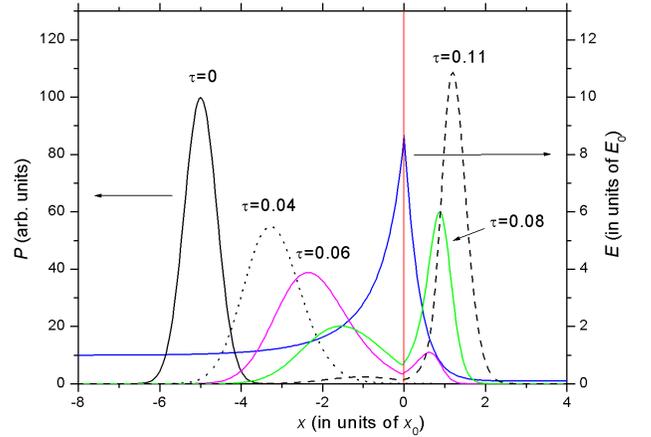

**Figure 2.** Dynamics of propagation through the boundary, of spin-polarized electrons injected at $\tau = 0$, where $\tau = t/\tau_{sf}$, for $N_2/N_1 = 10$. The blue curve denotes the electric field. The other curves show the spin polarization density at several times. Here $E_0 = j_0k_BT/(DN_1e^2)$, $(x_0)^{-2} = e^2N_1/(k_BT\varepsilon\varepsilon_0)$.

Evolution of the spin polarization density created at $t=0$ by the instantaneous source is shown in Figure 2. We tooked the profile of the initial spin polarization in the Gaussian form and solved Eq. (5) using an iterative scheme. Under influence of the electric field, the spin polarization density distribution moves towards the boundary. Its width increases due to diffusion, and the amplitude decreases due to the combined action of the different spin relaxation mechanisms and diffusion processes. As the spin polarization density profile approaches the boundary, its velocity increases. It reaches the maximum at the boundary, where the electric field is maximal. In the region with higher donor density, $N_2$, the electric field is lower, and the velocity of the spin polarization profile decreases. As a result, the spin polarization gathers in a narrow spatial region; see Figure 2.

The spin polarization at $\tau = 0.11$, in Figure 2, still represents a dynamical solution of Eq. (5). However, at such late times, the velocity of the spin polarization profile is much lower, at least by the factor $N_1/N_2$, than at earlier times, when the spin polarization density was concentrated mainly in the first semiconductor or in the interface region. Thus, the peaked spin polarization profile at $\tau=0.11$ can be regarded as quasi-static, its position and amplitude slowly varying in time. After a long time, it will dissipate due to spin relaxation and diffusion processes. Equation (5) was solved with the continuous source of spin polarization for different values of the doping density $N_2$. If the doping densities are equal ($N_1=N_2$), then the spin polarization density decreases monotonically with distance from the injection point. The peak value of the spin polarization density, past the boundary, increases as the doping density $N_2$ increases [19].

Physically, the mechanism of the spin polarization density amplification near the boundary at which the doping is changed, can be understood as follows. The spin polarization density can be increased near the boundary due to charge localization: the density of the electrons must be large in the $N_2$ semiconductor region. The electrons moving fast in the $N_1$ region, then move slowly in the $N_2$ region and gather in a small spatial region near the boundary.

In conclusion, the electron spin transport through the boundary between two semiconductor regions with different doping levels can lead to the electron spin polarization amplification near the boundary. The built-in electric field at the boundary accelerates propagation of the spin polarization through the boundary, if spin polarization passes from the low doping region to the high doping region. Spin amplification occurs past the boundary, within the distance of the order of the depletion layer width. It must be emphasized that there exists other mechanisms allowing increasing spin polarization density near a boundary. For example, the two semiconductor regions could have different diffusion coefficients; for a more efficient spin focusing near the boundary, a lower diffusion coefficient in the $N_2$ region would be desirable. However, as mentioned in the introduction, the mechanism involving only the doping variation, has the advantage of not requiring a materials interface, thus avoiding additional spin-polarization losses.

## ACKNOWLEDGMENTS

We gratefully acknowledge helpful discussions with Professors M.-C. Cheng, V. N. Gorshkov and I. D. Vagner.